\documentstyle[aps,twocolumn]{revtex}

\begin{document}
\twocolumn[\hsize\textwidth\columnwidth\hsize\csname@twocolumnfalse\endcsname

\title
{Possible realization of Josephson charge qubits in two coupled Bose-Einstein condensates}

\author{Zeng-Bing Chen$^{1}$ and Yong-De Zhang$^{2,1}$}
\address
{$^1$Department of Modern Physics, University of Science and Technology of 
China, Hefei, Anhui 230026, P.R. China}
\address
{$^2$CCAST (World Laboratory), P.O. Box 8730, Beijing 100080, P.R. China}
\date{\today}
\maketitle 

\begin{abstract}
\

We demonstrate that two coupled Bose-Einstein condensates (BEC) at zero
temperature can be used to realize a qubit which is the counterpart 
of Josephson charge qubits. The two BEC are weakly coupled and confined 
in an asymmetric double-well trap. When the ``charging energy" of the system 
is much larger than the Josephson energy and the system is biased 
near a degeneracy point, the two BEC represent a qubit with two states 
differing only by one atom. The realization of the BEC qubits in realistic 
BEC experiments is briefly discussed.
\

\


\

\
\end{abstract}]

Bose-Einstein condensates (BEC) in dilute systems of trapped neutral atoms 
\cite{BEC,RMP} have offered a new fascinating testing ground for some basic
concepts in elementary quantum mechanics and quantum many-body theory as
well as for searching new macroscopic quantum coherent phenomena. In
particular, much interest has been focused on the possibility of coherent
atomic tunneling between two trapped BEC \cite
{Java,Smerzi,Leggett,laser-assist,Chen-MPLB,Leggett-RMP}, which is analogous
to the superconducting Josephson effect.

Recent years have witnessed the rapid development of quantum information
theory. One of the central results there is the possibility of quantum
computation \cite{quan-inf,Nature}. In quantum computers, an elementary
building block is a set of two-level systems (``qubits'') storing quantum
information. Due to the superposition of quantum information states in
qubits, a quantum computer can provide fundamentally more powerful
computational ability than its classical counterpart. So far, many systems
have been proposed to perform quantum logic, including atoms in traps \cite
{CZ}, cavity quantum-electrodynamical systems \cite{CQED}, the nuclear
magnetic resonance system \cite{NMR}, and solid state systems (e.g., quantum
dots \cite{dot}, nuclear spins of atoms in doped silicon devices \cite{Kane}%
, and ultrasmall Josephson junctions \cite
{Shnirman,Averin,Makhlin,Ioffe,eprint}). Recently it has been demonstrated
that a single Cooper pair box is a macroscopic two-level system that can be
coherently controlled \cite{box-PRL,box}, thus realizing a {\it Josephson
charge qubit} \cite{Shnirman,Averin,Makhlin,eprint}. Making use of
superconducting quantum interference loops containing ultrasmall Josephson
junctions, the {\it Josephson flux qubit} has also been proposed \cite
{Ioffe,eprint,Mooij}. Two recent experiments indeed realized the quantum
superposition of macroscopic persistent-current states, and demonstrated the
small Josephson junction loops as macroscopic quantum two-level systems \cite
{JJcat}.

The implementation of qubits in a physical systems is the first step for
quantum computing. In this paper we demonstrate that two coupled BEC at zero
temperature can be used to realize a qubit. This BEC qubit is the
counterpart of the Josephson charge qubits. The two BEC are weakly linked
and confined in an asymmetric double-well trap. Typically, the system may
display the phenomenon known as quantum coherent tunneling of atoms, or
Josephson effect \cite{Java,Smerzi,Leggett,Chen-MPLB}. However, under the
conditions that the ``charging energy'' of the system is much larger than
the Josephson energy and the system is biased near a degeneracy point, the
two BEC may represent a qubit with two states differing only by one atom.

The effective many-body Hamiltonian describing atomic BEC in a double-well
trapping potential, $V_{trap}({\bf r})$, can be written in the
second-quantization form as 
\begin{eqnarray}
\hat H_{BEC} &=&\int d^3r\hat \psi ^{\dagger }\left[ -\frac{\hbar ^2}{2m}%
\nabla ^2+V_{trap}\right] \hat \psi  \nonumber \\
&&\ +\frac{U_0}2\int d^3r\hat \psi ^{\dagger }\hat \psi ^{\dagger }\hat \psi 
\hat \psi ,  \label{ham}
\end{eqnarray}
where $m$ is the atomic mass, $U_0=4\pi a\hbar ^2/m$ (with $a$ denoting the $%
s$-wave scattering length) measures the strength of the two-body
interaction, and the Heisenberg atomic field operators $\hat \psi ^{\dagger
}({\bf r},t)$ and $\hat \psi ({\bf r},t)$ satisfy the standard bosonic
commutation relation $[\hat \psi ({\bf r},t),\hat \psi ^{\dagger }({\bf r}%
^{\prime },t)]=\delta ({\bf r}-{\bf r}^{\prime })$. One can properly choose
the trapping potential so that the two lowest states are closely spaced and
well separated from the higher energy levels, and that many-body
interactions only produce negligibly small modifications of the ground-state
properties of the individual wells \cite{two-mode}. The potential has two
minima at ${\bf r}_1$ and ${\bf r}_2$, and can be expanded around each
minimum as 
\begin{eqnarray}
V_{trap}({\bf r}) &=&\sum_{i=1,2}\left[ V_i^{(0)}({\bf r}_i)+V_i^{(2)}({\bf r%
}-{\bf r}_i)\right] +\cdots  \nonumber  \label{trap} \\
&\equiv &V_1+V_2+\cdots ,  \label{trap}
\end{eqnarray}
where the two constant potentials $V_i^{(0)}({\bf r}_i)$ can be set to zero
without loss of generality.

With the above assumptions, one can use a two-mode approximation to the
many-body description of the system \cite{two-mode}. Thus instead of the
standard mode expansion over single-particle states, one can approximately
expand the field operators $\hat \psi ({\bf r},t)$ in terms of two local
modes \cite{Smerzi,two-mode} 
\begin{equation}
\hat \psi ({\bf r},t)\approx \sum_{i=1,2}\hat a_i(t)\Psi _i({\bf r}-{\bf r}%
_i),  \label{expan}
\end{equation}
where $[\hat a_i(t),\hat a_j^{\dagger }(t)]=\delta _{ij}$, and $\Psi _i({\bf %
r}-{\bf r}_i)\equiv \Psi _i({\bf r})$ are two local mode functions (with
energies $E_i^0$) of the individual wells and satisfy 
\begin{equation}
\int d^3r\Psi _i^{*}({\bf r})\Psi _j({\bf r})\approx \delta _{ij}.
\label{ortho}
\end{equation}
Substituting Eq. (\ref{expan}) into the effective Hamiltonian (\ref{ham})
yields the two-mode approximation of $\hat H_{BEC}$: 
\begin{eqnarray}
\hat H &=&\sum_{i=1,2}(E_i^0\hat a_i^{\dagger }\hat a_i+\lambda _i\hat a%
_i^{\dagger }\hat a_i^{\dagger }\hat a_i\hat a_i)  \nonumber \\
&&\ \ -(J\hat a_1^{\dagger }\hat a_2+J^{*}\hat a_1\hat a_2^{\dagger }),
\label{h2}
\end{eqnarray}
where the parameters are estimated by \cite{Smerzi,two-mode} 
\begin{eqnarray}
E_i^0 &=&\int d^3r\Psi _i^{*}\left[ -\frac{\hbar ^2}{2m}\nabla
^2+V_{trap}\right] \Psi _i,  \label{e7} \\
\lambda _i &=&\frac{U_0}2\int d^3r\left| \Psi _i\right| ^4,  \label{e8} \\
J &=&-\int d^3r\left[ \frac{\hbar ^2}{2m}{\bf \nabla }\Psi _1^{*}\cdot {\bf %
\nabla }\Psi _2+V_{trap}\Psi _1^{*}\Psi _2\right] .  \label{e9}
\end{eqnarray}
Without loss of generality, $J$ can be regarded a real number. Notice that
interactions between atoms in different wells are neglected in Eq. (\ref{h2}%
) \cite{two-mode}.

Defining two local number operators $\hat n_i=\hat a_i^{\dagger }\hat a_i$,
then it is easy to verify that the total number operator $\hat a_1^{\dagger }%
\hat a_1+\hat a_2^{\dagger }\hat a_2=\hat N$ of atoms represents a
conservative quantity and thus $\hat N=N_t\equiv 2N=const$. After neglecting
the constant term, the two-mode Hamiltonian $\hat H$ can be rewritten as 
\begin{eqnarray}
\hat H^{\prime } &=&E_c(\hat n-n_g)^2-J(\hat a_1^{\dagger }\hat a_2+\hat a_1%
\hat a_2^{\dagger }),  \nonumber \\
\hat n &\equiv &(\hat n_1-\hat n_2)/2,\;\;E_c=\lambda _1+\lambda _2, 
\nonumber \\
n_g &=&\frac 1{2E_c}[(E_2^0-E_1^0)+(N_t-1)(\lambda _2-\lambda _1)],
\label{h22}
\end{eqnarray}
where $\hat n$ is the number difference operator, $E_c$ is the ``charging
energy'' and $n_g$ acts as a control parameter and is known as the ``gate
charge'' in other context \cite{eprint}. Due to the conservation of total
atom number, the problem can be restricted to the subspace of definite $N$,
i.e., 
\begin{eqnarray}
\left| n_1,n_2\right\rangle &=&\left| N+n,N-n\right\rangle \equiv \left|
n\right\rangle ,  \nonumber \\
\hat N\left| n\right\rangle &=&2N\left| n\right\rangle ,\;\;\hat n\left|
n\right\rangle =n\left| n\right\rangle ,  \label{basis}
\end{eqnarray}
where $\left| n_1,n_2\right\rangle =\left| n_1\right\rangle _1\left|
n_2\right\rangle _2$, with $\left| n_i\right\rangle _i$ denoting the usual
number states of the two local modes ($\hat n_i\left| n_i\right\rangle
_i=n_i\left| n_i\right\rangle _i$). Using the basis $\left| n\right\rangle $
($n=N$, $N-1$, $\ldots $, $-N$), one easily obtains 
\begin{equation}
\hat a_1^{\dagger }\hat a_2\left| n\right\rangle =\sqrt{(N+n+1)(N-n)}\left|
n+1\right\rangle .  \label{aa}
\end{equation}
Consequently, $\hat a_1^{\dagger }\hat a_2$ in this subspace permits the
following representation in the basis: 
\begin{equation}
\hat a_1^{\dagger }\hat a_2=\sum_n\sqrt{(N+n+1)(N-n)}\left| n+1\right\rangle
\left\langle n\right| .  \label{aa1}
\end{equation}
Thus the Hamiltonian in Eq. (\ref{h22}) takes the form 
\begin{eqnarray}
\hat H^{\prime } &=&E_c(\hat n-n_g)^2-J\sum_n\sqrt{(N+n+1)(N-n)}  \nonumber
\\
&&\times \left[ \left| n+1\right\rangle \left\langle n\right| +\left|
n\right\rangle \left\langle n+1\right| \right] .  \label{hnn}
\end{eqnarray}

To realize the Josephson charge qubits in a small Josephson junction system
(a ``single-Cooper-pair box'') \cite{Shnirman,box-PRL}, two conditions
should be met: ($1$) The charging energy of the single-Cooper-pair box is
much larger than the Josephson energy; and ($2$) The system is biased near a
degeneracy point so that the system represents a qubit with two states
differing only by one Cooper-pair charge. Now suppose that two similar
conditions are also met in the present context. Then the Hamiltonian in Eq. (%
\ref{hnn}) takes the form 
\begin{eqnarray}
\hat H^{\prime } &=&E_c(\hat n-n_g)^2  \nonumber \\
&&\ -\frac{E_J}2\sum_n\left[ \left| n+1\right\rangle \left\langle n\right|
+\left| n\right\rangle \left\langle n+1\right| \right] ,  \label{hn}
\end{eqnarray}
in which $E_J\equiv 2J\sqrt{N(N+1)}\approx N_tJ$ is the Josephson energy. In
deriving Eq. (\ref{hn}), we have simplified Eq. (\ref{aa1}) further by 
\begin{equation}
\hat a_1^{\dagger }\hat a_2\approx \sqrt{N(N+1)}\sum_n\left|
n+1\right\rangle \left\langle n\right|  \label{nn1}
\end{equation}
if $N\gg \left| n\right| $ for most of the relevant number states of the
system. Physically, the operators $\left| n+1\right\rangle \left\langle
n\right| $ and $\left| n\right\rangle \left\langle n+1\right| $ transfer a
single atom between the two wells.

The form of $\hat H^{\prime }$ in Eq. (\ref{hn}) resembles the Hamiltonian
that is used to realize the Josephson charge qubits \cite{Shnirman,box-PRL}.
Similarly, one can introduce the phase-difference operator $\hat \varphi $
that is conjugate with $\hat n$: $\frac 12\left[ \left| n+1\right\rangle
\left\langle n\right| +\left| n\right\rangle \left\langle n+1\right| \right]
\equiv \cos \hat \varphi $. In this case, we recover the Hamiltonian widely
used in the literature \cite{eprint}: 
\[
\hat H^{\prime }=E_c(\hat n-n_g)^2-E_J\cos \hat \varphi . 
\]
However, it should be noted that there are some subtleties in doing so,
mainly in the context of elementary quantum mechanics, as is well known \cite
{phase,Chen-MPLB}. Yet, identifying $\hat n$ and $\hat \varphi $ as a
canonically conjugate pair is very popular and seems to be well justified in
the context of condensed matter physics, typically dealing with a large
number of particles \cite{Anderson,Leggett-RMP}.

The first condition stated above implies $E_c\gg E_J$. In this case, the
energies of the system's states are dominated by the charging part of the
Hamiltonian (\ref{hn}). The second condition means that when $n_g$ is
approximately half-integer (say $n_{\deg }+\frac 12$), the charging energies
of two adjacent states ($\left| n_{\deg }\right\rangle $ and $\left| n_{\deg
}+1\right\rangle $) are nearly degenerate, and as such the Josephson term
mixes the two adjacent states strongly. As a result, the eigenstates of the
Hamiltonian (\ref{hn}) are now superpositions of $\left| n_{\deg
}\right\rangle $ and $\left| n_{\deg }+1\right\rangle $ with a minimum
energy gap $E_J$ between them. In this case, the two coupled BEC form a BEC
qubit that is analogous to the Josephson charge qubits of ultrasmall
Josephson junctions. Thus the system under study acts as a ``single-atom
box'', a BEC counterpart of the single-Cooper-pair box \cite{box}. Similarly
to the case of ultrasmall superconducting Josephson junctions \cite
{Shnirman,eprint}, $\left| n_{\deg }\right\rangle \equiv \left| \uparrow
\right\rangle $ and $\left| n_{\deg }+1\right\rangle \equiv \left|
\downarrow \right\rangle $. The effective Hamiltonian in the spin-$\frac 12$
notion is correspondingly 
\begin{equation}
{\cal H}=-\frac{E_c^{\prime }}2\sigma _z-\frac{E_J}2\sigma _x.  \label{spin}
\end{equation}
Here 
\begin{eqnarray}
\sigma _z &=&\left| \uparrow \right\rangle \left\langle \uparrow \right|
-\left| \downarrow \right\rangle \left\langle \downarrow \right| ,\;\;\sigma
_x=\left| \uparrow \right\rangle \left\langle \downarrow \right| +\left|
\downarrow \right\rangle \left\langle \uparrow \right| ,  \nonumber \\
E_c^{\prime } &=&E_c[1-2(n_g-n_{\deg })],\;\;(n_g-n_{\deg })\sim 1/2,
\label{segma}
\end{eqnarray}
from which the dependence of ${\cal H}$ (or $E_c^{\prime }$) on the gate
charge $n_g$ and the degenerate point is evident. The expression of $%
E_c^{\prime }$ shows that though $E_c\gg E_J$, the $E_J$-term may dominate
the $E_c^{\prime }$-term near the degenerate point. With the Hamiltonian (%
\ref{spin}) at hand, the single-qubit operations can be achieved for the BEC
qubit.

To this end, it is important to notice that in realistic BEC experiments,
the double-well potential can be created by using a far-off-resonance
optical dipole force to perturb a magnetic-rf trap \cite{Andrews}. As noted
in Ref. \cite{Smerzi}, the population of atoms in the individual well can be
monitored by phase-contrast microscopy; the shape of the double-well
potential can be tailored by the position and the intensity of the laser
sheet partitioning the magnetic trap. Thus it is experimentally feasible to
control the charging energy $E_c$, the Josephson energy $E_J$ and the
degenerate point. With the above remarks in mind, one can see that the BEC
qubit may be manipulated as the superconducting Josephson charge qubits \cite
{Shnirman}. For practically realizing the BEC qubits, one needs to see
whether the condition $E_c\gg E_J$ can be met in realistic BEC experiments.
The condition is usually known as the Fock regime, which is attainable by
adjusting $E_c$ and $E_J$ for the system under study \cite
{Leggett,Leggett-RMP}.

On the other hand, we would like to mention the difficulty of measuring the
BEC qubits. However, due to the beautiful technique of loading and detecting
individual atoms developed recently \cite{single-atom}, it is possible to
overcome this difficulty in a near future such that one can perform
experiments on the BEC qubits in which the single atoms can be detected with
sufficient efficiency.

Finally, it is worth pointing out that the same (or similar) Hamiltonian as
in (\ref{h2}) also arises in other context, e.g., in a two-mode nonlinear
optical directional coupler \cite{opt-2m} and in two-species BEC \cite
{BEC-cat}. The latter has been proposed to create macroscopic quantum
superpositions (the ``Schr\"odinger cat states''). Thus the present work may
be viewed in a wider context.

In conclusion, we have suggested that two coupled BEC at zero temperature
can be used to realize the BEC counterpart of the superconducting Josephson
charge qubits under the conditions that the charging energy of the system is
much larger than the Josephson energy and the system is biased near a
degeneracy point. If the conditions can be met, the two BEC may represent a
single-atom box, with two states differing only by one atom. It remains to
be seen how non-zero temperatures affect the predicted BEC qubits. Due to
extremely good coherence of BEC observed so far, we expect decoherence would
not render the BEC qubits unobservable. Another interesting issue is how to
entangle two BEC qubits.

Note added--Yu Shi recently brought our attention to his related work \cite
{Yu-Shi}.

We are grateful to the Referee for valuable suggestion. This work was
supported by the National Natural Science Foundation of China, the Chinese
Academy of Sciences, and by the Laboratory of Quantum Communication and
Quantum Computation of USTC.

\end{document}